\documentclass[global,twocolumn]{svjour}
\usepackage{latexsym}
\usepackage{graphics}
\usepackage{graphicx}
\usepackage{amsmath}
\usepackage{amssymb}
\usepackage{bm}

\journalname{Appl. Phys. B}
\begin{document}
\title{All-optical formation of a Bose-Einstein condensate for applications in scanning electron microscopy}
\author{T. Gericke\inst{1} \and P. W{\"u}rtz\inst{1}\and D. Reitz\inst{1}\and C. Utfeld\inst{2}\and H. Ott\inst{1}}
%
%
\institute{Institut f{\"u}r Physik, Johannes Gutenberg-Universit{\"a}t, 55099 Mainz, Germany 
			\and H.\,H.\,Wills Physics Laboratory, University of Bristol, Tyndall Avenue, Bristol BS8 1TL, UK}
\mail{e-mail: ott@uni-mainz.de}
\date{Received: date / Revised version: date}

\maketitle
\begin{abstract}
We report on the production of a F\,=\,1 spinor condensate of $^{87}$Rb atoms in a single beam optical dipole trap formed by a focused CO$_2$ laser. The condensate is produced $13\,\textrm{mm}$ below the tip of a scanning electron microscope employing standard all-optical techniques. The condensate fraction contains up to 100,000 atoms and we achieve a duty cycle of less than 10\,s. 
\end{abstract}

\section{Introduction}
\label{intro}

Since the first observation of Bose-Einstein condensation, ultracold atoms have proven to be an extremely rich system to study fundamental quantum effects \cite{Cornell02,Ketterle02}. Most of the experiments use absorption imaging in combination with time of flight techniques to analyze the system. The performance of this imaging technique is subjected to two fundamental limitations. The best possible spatial resolution is comparable to the wavelength of the absorption laser and the technique is not sensitive to single atoms. We are developing a new imaging technique which is able to detect single atoms inside a quantum gas with high spatial resolution. The technique is based on the principles of scanning electron microscopy. Atoms inside the atomic cloud are ionized by a focused electron beam and subsequently detected with high efficiency on an ion detector \cite{Gericke06}. This method overcomes both limitations. First, the spatial resolution is set by the diameter of the focused electron beam which can be made almost arbitrarily small. Second, the technique is by its nature capable to detect single atoms with high signal to noise ratio. With a resolution of better than 200\,nm the typical interatomic distances inside a quantum gas can be resolved and new measurements of atomic correlations become possible. For applications in optical lattices this technique offers the intriguing perspective to address the atoms in individual lattice sites. 

Because the electron beam would be strongly distorted by an inhomogeneous magnetic trapping field, we have chosen an all-optical approach for the production of the quantum gas \cite{Barrett01,Granade02,Cennini03,Jochim03,Takasu03,Kinoshita05}. Besides the high efficiency in loading and evaporating atoms in optical dipole traps and the possibility of employing Feshbach resonances in order to manipulate the interatomic interaction, all-optical approaches also offer a favorable alternative whenever experimental issues complicate the use of magnetic trapping potentials. The required optical access can be achie\-ved with relatively moderate experimental efforts, while the general advantage of fast evaporation and resulting short experimental cycle periods can be fully exploited. Especially the use of a single beam CO$_2$ laser dipole trap has proven to be a simple, robust and easy technique for the production of degenerate quantum gases \cite{Cennini03}.

In this work, we present experimental data of the production of a $^{87}$Rb Bose-Einstein condensate (BEC) inside a scanning electron microscope. We use a single beam optical dipole trap formed by a focused CO$_2$ laser beam and achieve Bose-Einstein condensation with high atom number and short experimental cycle time. Special emphasis is given to the particularities of our apparatus and to the compatibility of the experimental environment for the production of ultracold quantum gases and scanning electron microscopy.

\section{Experimental Setup}

\begin{figure}[h,t,b]
	\includegraphics[width=0.48\textwidth]{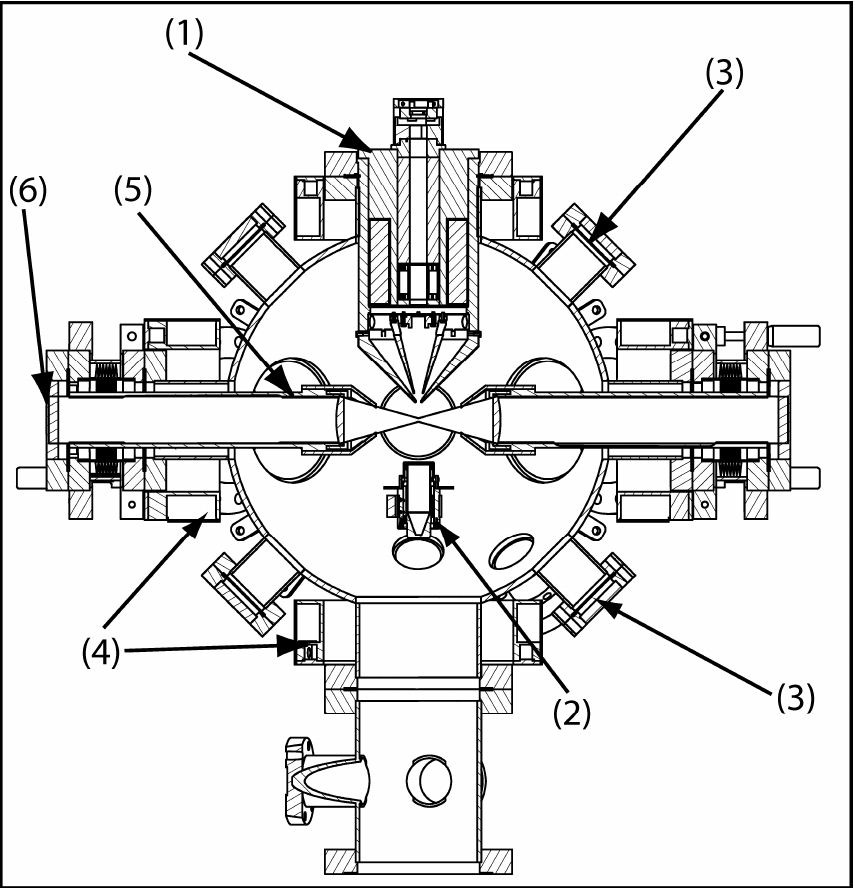}
\caption{Experimental setup, overview: electron column (1), Faraday cup (2), viewports for the MOT beams (3), coils (4), holders for the ZnSe lenses (5), ZnSe viewports (6). The spherical vacuum chamber has a diameter of $30\,$cm. See also Fig.\,\ref{fig2} for more details.}
\label{fig1}       
\end{figure}

The guidelines for the design of our apparatus were experimental simplicity concerning the production of the BEC and technical compatibility concerning the implementation of the electron column. The setup is based on a spherical vacuum chamber (main chamber) with a diameter of $30\,\textrm{cm}$ (see Fig.\ref{fig1}). The electron column is mounted on top of the main chamber. The end part of the column has a conical shape and the opening for the electron beam is located 15\,mm above the center of the chamber. In order to dump the electron beam and reduce the amount of secondary and backscattered electrons a Faraday cup is installed five centimeters underneath the tip of the electron column. The electron beam can be adjusted with aid of a test target which is fixated on a movable stainless steel rod. The optical dipole trap is formed by a CO$_2$ laser beam which intersects the main chamber in the horizontal direction. For this purpose, two ZnSe viewports are installed opposite to each other. The laser beam is focused (and collimated) with two ZnSe lenses mounted inside the vacuum on holders which are welded on flexible bellows. With three micrometer screws in a kinematic mount arrangement, the position of the two lenses can be adjusted independently in all three directions. For the future detection of the created ions simple ion extraction optics and a channeltron detector are mounted in the horizontal plane under $45^{\circ}$ with respect to the dipole trap axis. The main chamber is pumped by a 150\,l/s ion pump in combination with a titanium sublimation pump. A careful choice of materials used -- especially for the electron column \cite{column} -- guarantees a vacuum level of $1\times10^{-10}\,$mbar. The main chamber is encased with a 2\,mm thick $\mu$-metal shielding in order to reduce the influence of magnetic stray fields on the electron beam.

We use $^{87}$Rb atoms which are loaded from a 2D-magneto optical trap (2D-MOT) into a six-beam MOT. The vacuum chamber for the 2D-MOT is made out of a titanium cuboid with two orthogonal elongated through boreholes with a diameter of $25\,\textrm{mm}$ and a length of $81\,\textrm{mm}$. This provides a cylindrical cooling volume with a large aspect ratio. The windows are glued to the chamber to ensure a compact design. An oven with a rubidium reservoir is used as atom source. Four rectangular coils generate a two dimensional magnetic quadrupole field with a gradient of $19\,\textrm{G}/\textrm{cm}$. The cooling radiation is obtained by a tapered amplifier which is $16\,\textrm{MHz}$ red-detuned to the F\,=\,2$\,\rightarrow$\,F$\textrm{'}$\,=\,3 transition of the $\textrm{D}_2$ line. The F\,=\,1$\,\rightarrow$\,F$\textrm{'}$\,=\,2 repumping light of the $\textrm{D}_2$ line is generated by a grating stabilized diode laser. We combine $150\,\textrm{mW}$ of cooling and $5.5\,\textrm{mW}$ of repumping laser light on a polarization beam splitter cube and enlarge the beam with a telescope to a diameter of $25\,\textrm{mm}$. We mimic cylindrical beams for the laser cooling by using three spherical parallel beams instead. Each beam is then individually retroreflected. To enhance the emerging atom flux we superimpose a resonant push beam in the longitudinal direction of the quadrupole field. With this configuration we achieve a flux of up to $4\times 10^8\,\textrm{atoms}/{\textrm{s}}$ of $^{87}\textrm{Rb}$ at a rubidium partial pressure of about $4\times 10^{-7}\,$ mbar. The 2D-MOT chamber is connected to the main chamber via a differential pumping unit and a flexible bellow. We do not find any significant influence on the vacuum level of the main chamber. 

The MOT is located at the center of the main chamber and the strong axis of the magnetic quadrupole field lies in the horizontal plane perpendicular to the optical dipole trap axis. This configuration is favorable with regard to a maximum spatial overlap between the MOT and the optical dipole trap. The lower part of the electron column is made of an iron-nickel alloy with high magnetic permeability and thus any external magnetic quadrupole field is distorted. We circumvent this problem by using in total six coils which are oriented perpendicular to each other around the chamber (Fig.\,\ref{fig1}). The combined field of all six coils creates a spherical magnetic quadrupole field with the desired geometry and a volume of about $1\,\textrm{cm}^3$ \cite{volume}. The field minimum is located $2\,\textrm{mm}$ above the center of the chamber and the field gradient is $10\,\textrm{G}/\textrm{cm}$ along the strong axis. The laser light is obtained from the same laser system as described above. We use $120\,\textrm{mW}$ ($5.3\,\textrm{mW}$) for the cooling (repumping) light. Both are combined on a polarizing beam splitter and then divided into 6 beams with a diameter of 25\,mm each. An acousto-optical modulator (AOM) is implemented in the optical path for the repumping light. Precise control of its intensity is crucial for an efficient loading of the CO$_2$ dipole trap.

\begin{figure}[h,t,b]
	\includegraphics[width=0.48\textwidth]{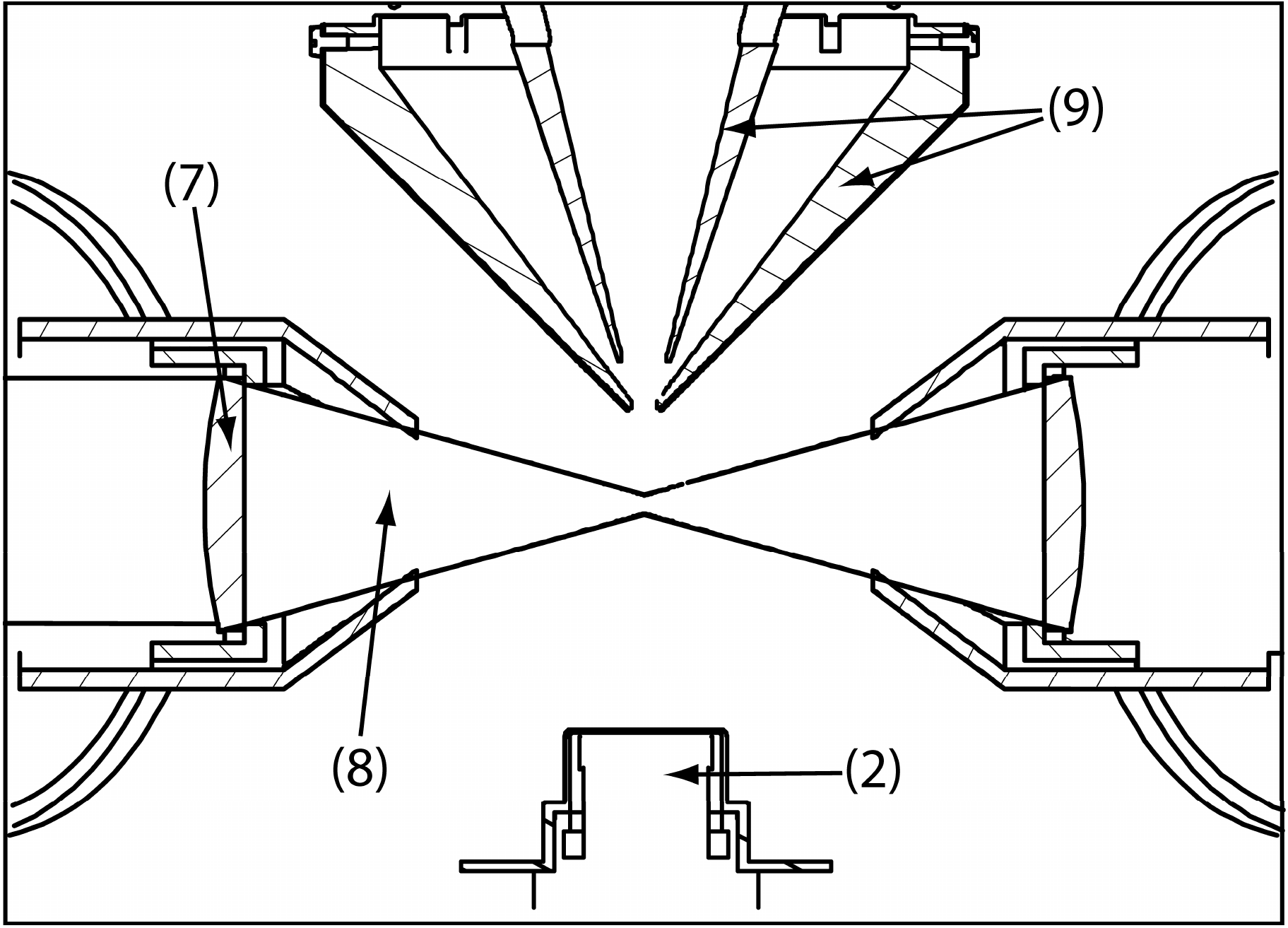}
\caption{Experimental setup, detailed view: Mounted ZnSe lenses (7), Faraday cup (2), sketched optical dipole trap (8), pole pieces of the magnetic lens (9). The ion detector (not shown) is mounted in the horizontal plane under $45^{\circ}$ with respect to the dipole trap axis.}
\label{fig2} 
\end{figure}

The optical dipole trap is formed by a focused CO$_2$ laser beam ($\lambda = 10.6\,\mu \textrm{m}$). We use an asphercial ZnSe lens ($f=63\,$mm) which provides a diffraction limited beam waist of $w_0=30\,\mu\textrm{m}$. A second lens is used to collimate the beam and guide it outside the vacuum chamber. The intensity of the CO$_2$ laser is controlled with an AOM. The AOM is driven with up to 30\,W of radio frequency (rf) power at a frequency between 30 and 50\,MHz. As some part of the rf power is absorbed inside the germanium crystal we observe strong thermal effects resulting in pointing instabilities. This constitutes a common problem in applications of CO$_2$ lasers \cite{Kobayashi06}. We strongly reduce this effect using two rf frequencies (35\,MHz and 45\,MHz) which are simultaneously applied to the AOM and whose combined power is constant \cite{Griffin06}. However, only one frequency (45\,MHz) fulfills the Bragg condition for efficient diffraction. Thus, the ratio of the power between these two frequencies determines the laser intensity in the dipole trap. With this method we achieve a pointing stability of better than 2\,mm at a distance of 2\,m behind the AOM. The focus of the CO$_2$ laser overlaps with the center of the MOT. At the maximum available power of 30\,W the depth of the dipole trap is $1.6\,$mK. Magnetic field gradients can limit the use of optical dipole traps because the magnetic force can easily expel atoms from the trap. In order to achieve a high current in the electron beam we have decided in favour of magnetic lenses for the electron optics \cite{magneticlens}. This puts strong requirements on the shielding of the magnetic fields in order to avoid large magnetic field gradients at the position of the atoms. The final magnetic lens which focuses the electron beam creates strong magnetic fields of up to 2000\,G only $2.5$\,cm above the center of the chamber. Therefore, we have carefully checked that the self shielding of the pole pieces (see Fig.\,\ref{fig2}) is sufficient to suppress magnetic fields leaking out of the tip to less than 1\,G at the position of the atoms.

\begin{figure}[h,t,b]
	\includegraphics[width=0.48\textwidth]{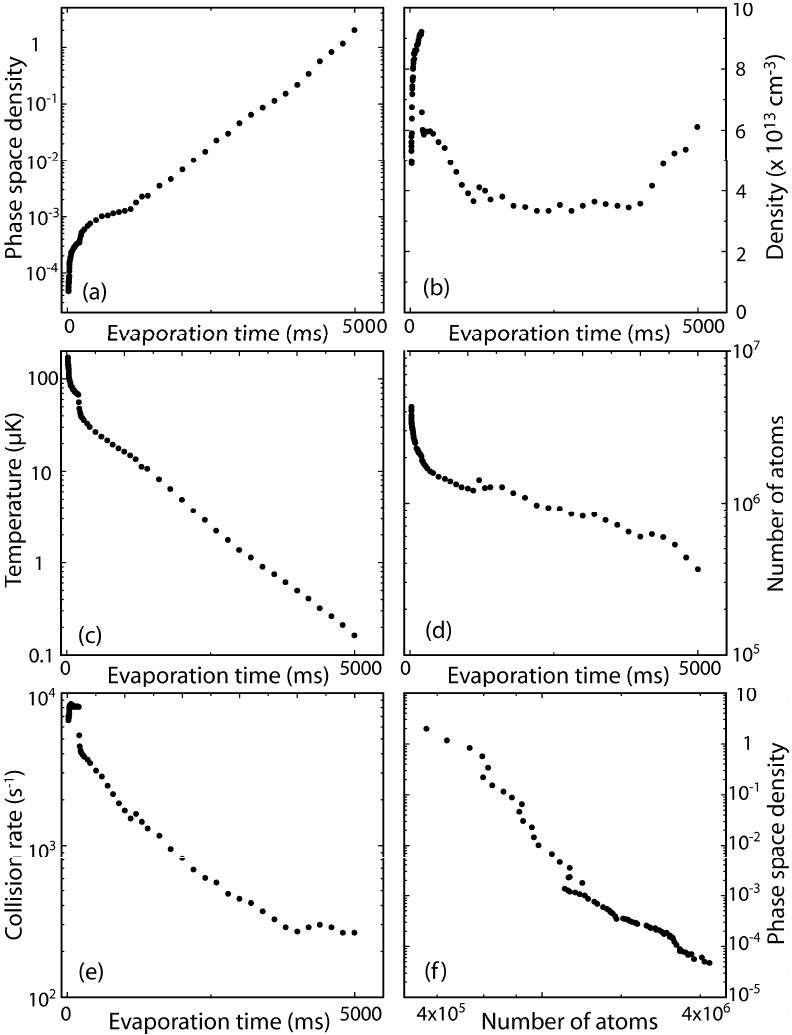}
\caption{Evolution of the phase space density (a), the density (b), the temperature (c), the number of atoms (d), and the collision rate (e) during the evaporation process. The last plot (f) shows the evolution of the phase space density versus the number of atoms (to be read from right to left). }
\label{fig3}       
\end{figure}

\section{Results}

At the beginning of every experimental cycle we load the MOT from the 2D-MOT for 3\,s. Within this time we collect $1\times10^9$ atoms in the MOT. The detuning of the MOT is equal to the detuning of the 2D-MOT ($16\,\textrm{MHz}$). The CO$_2$ laser is switched on during the MOT loading phase. After 3\,s we turn off the laser light of the 2D-MOT and the loading sequence of the dipole trap begins. We first set the detuning of the MOT to 25\,MHz for 40\,ms. Within 2\,ms we then rapidly increase the detuning in two steps to 35\,MHz and to 170\,MHz. At the same time we start to reduce the intensity of the repumping light linearly within 5\,ms to $1.6$\,\% of its original value. After 25\,ms we switch off the cooling and the repumping light. The magnetic quadrupole field as well as the intensity of the dipole trap laser are kept constant during the loading sequence. The best loading efficiency is achieved for an initial CO$_2$ laser power of 10\,W, corresponding to a trap depth of $530\,\mu$K. We load $4\times10^6$ atoms with an initial temperature of $170\,\mu$K into the dipole trap. The background gas limited lifetime of the atomic sample is 25\,s and the measured trap frequencies are $\omega _\textrm{ax}=2\pi\times 180$\,Hz ($\omega _r=2\pi\times 2860$\,Hz) in the axial (radial) direction. This results in a collision rate of $\gamma =6800\,$s$^{-1}$, a density of $n=5\times10^{13}\,$cm$^{-3}$ and a phase space density of $n\lambda_\mathrm{dB}^3=5\times 10^{-5}$, where $\lambda_\mathrm{dB}$ is the thermal de Broglie wavelength. We trap all three Zeeman substates of the F\,=\,1 manyfold. Therefore, the calculated phase space density includes a factor of $1/3$ and represents the phase space density of each individual spin state.  We leave the trap for $200\,\textrm{ms}$ at constant intensity allowing for plain evaporation. After this time the temperature has dropped to $70\,\mu \textrm{K}$ ($\gamma =8000\,$s$^{-1}$, $n=9\times10^{13}\,$cm$^{-3}$, and $n\lambda_\mathrm{dB}^3=3.5\times 10^{-4}$). Further evaporation is achieved by lowering the intensity of the CO$_2$ laser. We find the best results for a double exponential ramp. The time constants are $ 10\,\textrm{ms}$ for the fast decay (amplitude of $4\,\textrm{W}$) and $ 950\,\textrm{ms}$ for the slow decay (amplitude $6\,\textrm{W}$). After 5\,s of evaporation the atomic ensemble reaches the quantum degeneracy.

\begin{figure}[h,t,b]
\begin{center}
	\includegraphics[width=0.35\textwidth]{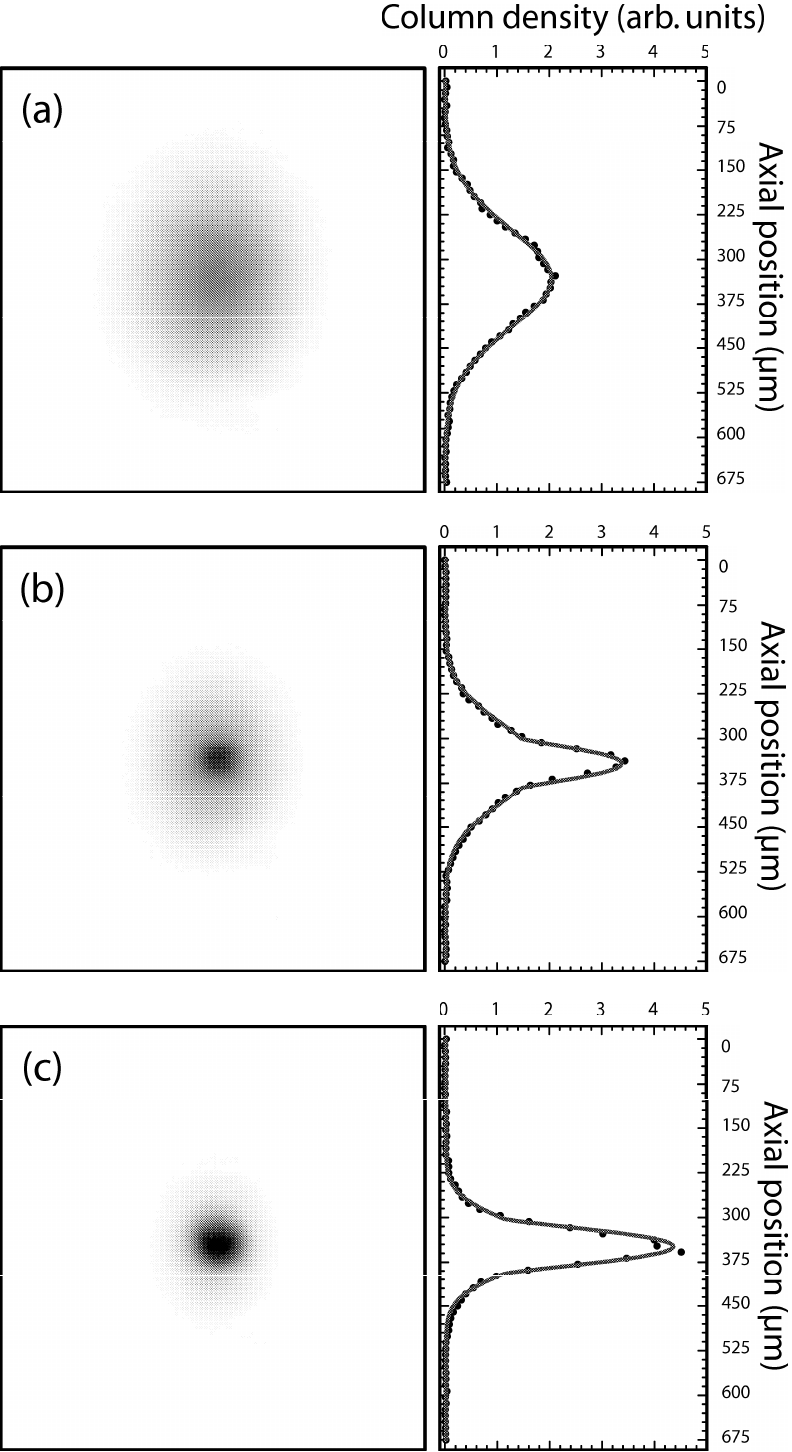}
\caption{Typical absorption images of the atomic cloud after 15\,ms time of flight for different temperatures. (a) Thermal cloud ($T=250\,$nK, $N_\textrm{th}=5\times10^{5}$). (b) Onset of Bose-Einstein condensation ($T=150\,$nK, $N_0=2.5\times10^{4}$, $N_\textrm{th}=2.8\times10^{5}$). (c) Bimodal distribution ($T=70\,$nK, $N_0=6\times10^{4}$, $N_\textrm{th}=1.4\times10^{5}$).}
\label{fig4}       
\end{center}
\end{figure}

In Fig.\,\ref{fig3} we show the evolution of the temperature, the number of atoms, the phase space density, the collisional rate and the density during the first 5\,s of the evaporation process. We reach the critical temperature for Bose-Einstein condensation ($T_c=160\,n$K) with $4\times10^5$ atoms. The central density at this point is $n=6\times10^{13}\,$cm$^{-3}$ and the collision rate in the trap center is $\gamma =250\,$s$^{-1}$. Losing one order of magnitude in atom number we gain nearly five orders of magnitude in phase space density (see Fig.\,\ref{fig3}f). This is close to the theoretical value for evaporation processes in a single beam CO$_2$ dipole trap as analyzed in \cite{Ohara01} and proves the very high efficiency of our evaporation process. Further reduction of the laser intensity results in a large F\,=\,1 spinor condensate. The final power in the optical dipole trap is $50\textrm{mW}$ and the measured trap frequencies are $\omega_{\textrm{ax}}=2\pi\times 12\,\textrm{Hz}$ in the axial and $\omega_{r}=2\pi\times 170\,\textrm{Hz}$ in the radial direction. Typical absorption images above, close to and below the critical temperature for Bose-Einstein condensation are shown in Fig.\,\ref{fig4}. The largest condensates that we can produce contain about 100,000 atoms.

\section{Conclusions}
\label{sec:3}
We have achieved all-optical Bose-Einstein condensation of $^{87}$Rb atoms. We use a single beam CO$_2$ laser dipole trap and produce in less than 10\,s condensates with up to 100,000 atoms. This is three times more than previously reported \cite{Freiburg07} and demonstrates impressively the power of this very simple technique. The condensate is produced inside a scanning electron microscope and the production scheme of the condensate is fully compatible with the technology of modern electron microscopes. The magnetic field management and shielding in combination with the use of an all-optical approach ensures the distortion free operation of the electron microscope, the MOT and the optical dipole trap. This technological breakthrough now paves the way for novel studies of degenerate quantum gases with unprecedented spatial resolution combined with single atom sensitivity.

%
\begin{acknowledgement}
We acknowledge the technical assistance of N. Hommerstad in the construction of the apparatus. We gratefully acknowledge financial support from the DFG under Grant No. Ot 222/2-3.
\end{acknowledgement}

\begin{thebibliography}{}
\bibitem{column}
The electron column uses magnetic lenses whose pole pieces are made of an iron nickel alloy. The surface of the alloy has been subjected to a special temperature treatment in order to achieve ultra high vacuum compatibility.
\bibitem{volume}
Within this volume, the magnetic field gradient grows linearly in each direction and resembles that of a spherical quadrupole field.
\end{thebibliography}
%

\end{document}